\documentclass[12pt]{spie} 
\usepackage{graphicx}
\usepackage{amssymb} 
\usepackage{amsmath}    
\usepackage{latexsym}   
\usepackage{bm}
\usepackage{times}

\def\abr{\mathrel{\stackrel{\longrightarrow}{A\!B}}\!}

\title{Irreversibility in physics stemming from unpredictable symbol-handling agents}

\author{John M. Myers\supit{a} and F. Hadi Madjid\supit{b}
\skiplinehalf
\supit{a}Harvard School of Engineering and Applied 
Sciences, Cambridge, MA 02138, USA \\
\supit{b}82 Powers Road, Concord, MA 01742, USA}

\authorinfo{Further author information: (Send correspondence to J.M.M.)\\J.M.M.: 
  F.H.M.: E-mail: gmadjid@aol.com, Telephone: 1 978 369 1808\\
E-mail: myers@seas.harvard.edu, Telephone: 1 617 495 5263}

\begin{document}
\maketitle

\begin{abstract}
The basic equations of physics involve a time variable $t$ and are invariant
under the transformation $t \rightarrow -t$. This invariance at first sight
appears to impose time reversibility as a principle of physics, in conflict
with thermodynamics. But equations written on the blackboard are not the whole
story in physics.  In prior work we sharpened a distinction obscured in today's
theoretical physics, the distinction between obtaining evidence from
experiments on the laboratory bench and explaining that evidence in
mathematical symbols on the blackboard.  The sharp distinction rests on a proof
within the mathematics of quantum theory that no amount of evidence,
represented in quantum theory in terms of probabilities, can uniquely determine
its explanation in terms of wave functions and linear operators.  Building on
the proof we show here a role in physics for unpredictable symbol-handling
agents acting both at the blackboard and at the workbench, communicating back
and forth by means of transmitted symbols.  Because of their unpredictability,
symbol-handling agents introduce a heretofore overlooked source of
irreversibility into physics, even when the equations they write on the
blackboard are invariant under $t \rightarrow -t$.  Widening the scope of
descriptions admissible to physics to include the agents and the symbols that
link theory to experiments opens up a new source of time-irreversibility in
physics.

\end{abstract}

\section{Introduction}
In 1950's, Sch\"odinger complained that the physics he helped make has no
place in it to represent himself \cite{schrodinger}.  These days Bayesians
view probabilities as numbers assigned by {\em agents}, and `Quantum
Bayesians' ascribe the choice of wave functions to agents \cite{fuchs2011}.
In quantum information science, Alice and Bob are spoken of as agents.  These
ways of speaking of agents are a start that we propose to advance by
recognizing structure necessary to the functioning of agents in a
communications network.

I want to think of myself as an agent to be represented in physics.  Not all
of me, but a slice of my activity.  Which slice? The slice in which something
I do might matter to physics.  Is there any such slice?  An answer to this
question came in 2005 with the proof within quantum theory that whatever
evidence one may have leaves open an infinite set of inequivalent
explanations, so that an investigator faces a choice undetermined by logic:
picking an explanation takes a guess \cite{ams02,aop05,CUP}.  Furthermore the
guess enters the design and operation of physical experiments, suggesting
that acts of guessing might be worth representing as acts of an agent
expressed within theoretical physics.  With this in mind, the idea came in
2002 of picturing myself as an investigator seated at the console of a
classical process-control computer (CPC) managing an experiment---programming
and running algorithms that control computer-mediated feedback loops, sending
commands to actuators on the workbench, recording occurrences of outcomes,
and modifying guessed hypotheses \cite{ams02}.

The design and operation of an experiment brings explanations written in
symbols on the blackboard into contact with the devices on the workbench. In
designing an experiment or in interpreting its results, an investigator faces
choices of explanations and of physical devices, and the investigator must
choose, and, on occasion, modify, which explanation to link to which
arrangement of devices. The activity of such an investigator is recordable in
files of a CPC used both to calculate with the equations and to manage the
instruments. By noticing that equations and instruments make contact in a
CPC, we imagine the traffic in symbols transmitted by the CPC to and from the
workbench and the blackboard or our own imagination as a heretofore
unrecognized part of physics.

On what basis can we argue for including traffic in transmitted symbols,
including numerals, as part of physics?  In special relativity Einstein
defined spacetime in terms of numerical clock readings at the transmission of
a signal from one clock and the reception of a signal at another clock
\cite{einstein05}.  Hence implicit in the concept of spacetime is the
propagation of signals that carry symbols, namely the numerals by which a
clock reading can be expressed \cite{qip15}.  On this basis we claim the
right to bring the propagation of signal-carrying symbols along with the
symbols carried into theoretical physics, and to introduce {\em agents} to
transmit and to receive symbols.  Here we introduce what we call a {\em
  symbol-handling agent} as a role played sometimes by a person and other
times by an automated device.  In developing a notion of symbol-handling
agents as elements of discourse in physics, we want to keep these agents
``lean.''  We begin with the idea that an agent makes unpredictable responses
to unpredictable inputs.  We limit ourselves as much as we can to this focus
on unpredictability, attributing to symbol-handling agents neither values nor
volition.  By keeping agents ``lean'' in this way we soften a boundary
between animate and inanimate to show a wealth of unsuspected structure,
relevant to both persons and to computer-based machinery as agents.  Next
comes the question: what difference to physics does it make to allow for
symbol-handling agents in physical descriptions?  An answer worked out below
is that the introduction of symbols such as numerals along with
symbol-handling agents brings a new source of irreversibility into physics.

In the next section we introduce a model of signal-handling agents, focused
on displaying an unsuspected structure of \emph{logical synchronization}
necessary for symbols transmitted by one agent to be recognizable to another
agent.  In Sec.\ \ref{sec:3} we show how unpredictable, symbol-handling
agents imply irreversibility.  In Sec.\ \ref{sec:4} we review the notion of
``agent'' that has been rejected in physics for three centuries but is now,
in modified form, attracting advocates.

\section{Model of a symbol-handling agent}\label{sec:2}
By \emph{symbols}, we mean recognizable elements of communication among
agents, human or not, that can lead to action involving energy supplied not
by the symbols themselves, nor by the agent that sends them, but by the
receiving agent.  (I can speak in words as symbols to ask you to move; if I
just push you, that is not communication by use of symbols.)  We conceive of
a symbol-handling agent, human or not, that computes with numerals,
communicates sequences of symbols, especially numerals, with other agents,
and takes unpredictable symbolic input and also interacts, energetically
rather than symbolically, with an environment.

We posit that a symbol-handling agent is equipped with a memory that at any
moment may be found in one of some number of states.  How a receiving agent
responds to a symbol is contingent on the contents of the memory of the
agent. Examples of symbols are (1) instances of letters, words, and numerals,
expressed as written characters, (2) electronic impulses carrying bits 0 and
1 in a computer, and (3) molecules and their constituent parts
(e.g. nucleobases of DNA) involved in biological signaling. Symbols used by
people convey elements of thought, often prompting actions and emotions.  And
beyond `elements of thought', symbols convey the calculational traffic to be
found both in man-made computers and in the biological processes of living
creatures.  We speak of a \emph{symbol} (as something that can be recognized)
carried by a {\em signal} (as something physical that can be measured). It is
important to distinguish be `recognizing a symbol' from `measuring a signal
that carries the symbol' \cite{1639}.  For example, the symbols 0 and 1 are
carried over transmission lines of a computer chip by electromagnetic
signals.  A programmer is indifferent to how such signals vary in their
electrical characteristics and their timing within allowed tolerances, while
an electrical circuit designer has to attend to the details of the physical
signals that carry what the programmer views as symbols.  In particular, the
circuit designer must provide for maintaining the signal within allowable
tolerances, which requires that the computer hardware measure and respond to
small timing variations to which the programmer is oblivious, but on which
the operation of the computer depends.  It is noteworthy that symbols carried
by signals with large tolerances can take part in mechanisms that act with
tight tolerances.  For example the electronic signals in a computer can vary
over a fairly wide range in amplitude and timing without spoiling the
exactness of arithmetical operations of the computer.

Picture agents at the blackboard reading and writing explanations in
mathematical symbols, and picture agents at the workbench manipulating
arrangements of physical devices. Communication of symbols between agents at
the workbench and agents at the blackboard links evidence to explanations. As
will be described later, agents maintain a structure of logical
synchronization with one another, needed for symbols to be recognized.  For
this the agents respond to unpredictable, measured phases of reception of
signals, which requires their measuring the signals, rather than just
recognizing the symbols.

In order to produce a more formal expression of a symbol-handling agent, we
turn to Turing's 1936 paper in which he introduced Turing machines, thereby
establishing computability as a discipline within mathematics.  In a side
remark, he briefly introduced an alternative machine called a {\em choice
  machine}, contrasted with the usual Turing machine that Turing called an
a-machine:
\begin{quote}
  If at each stage the motion of a machine \ldots is completely determined by
  the configuration, we shall call the machine an ``automatic machine'' (or
  a-machine). For some purposes we might use machines (choice machines or
  c-machines) whose motion is only partially determined by the configuration
  \dots. When such a machine reaches one of these ambiguous configurations, it
  cannot go on until some arbitrary choice has been made by an external
  operator. This would be the case if we were using machines to deal with
  axiomatic systems \cite{turing}.
\end{quote}
Our model of a symbol-handling agent amounts to a choice machine modified in
two ways.  The first modification provides for a choice machine not only to
receive symbols but to transmit then, so that modified choice machines can
communicate symbols with one another and with an external environment.  The
second modification provides the choice machine with a clock that steps it,
with the tick rate of the clock adjusted by commands issuing from the choice
machine itself.  This is our model of symbol-handling agent.  In this model,
a symbol-handling agent ticks like a clock, computes like a digital computer,
and transmits and receives sequences of symbols, in some cases unpredictably.
To assure the unpredictability of a symbol-handling agent, we posit that an
``external operator'' chooses a symbol and writes it onto the scanned square
of the agent's choice machine {\em privately}, in the sense that the symbol
remains unknown to other agents unless and until the symbol-handling agent
that receives the chosen symbol reports it to others.  In past work we
described how such symbol-handling agents, which in that work we called {\em live
  clocks}\cite{qip15} or {\em open machines}\cite{aop14}, generate evidence
on which implementations of spacetime coordinates depend.

\subsection{Logical synchronization}
One can describe the operation of a Turing machine as one might describe the
play of a chess game by stating the conditions of the chess pieces on the
board over a sequence of moments interspersed by the moves of the players;
indeed, Turing speaks moments interspersed by moves.  For the choice machines
adapted to communicate as symbol-handling agents, one must attend in more
detail to the cycle of operation that encompasses a single moment along with
a single move.  We speak of phases of the cycle, thought of as indicated by
an imagined clock hand that rotates during a cycle once around a dial. For a
symbol to be received by a symbol-handling agent, the signal carrying the
symbol must arrive within a suitable phase of the cyclical operation of the
agent.  Otherwise the agent ``drops the ball''. An agent $B$ receiving
symbols from an agent $A$ under this condition is said to be \emph{logically
  synchronized} to agent $A$.  As discussed in [\citenum{aop14,qip15}],
maintenance of logical synchronization requires feedback loops that respond
to the measured deviation of the arrival of a signal from a desired aiming
phase.  This measured deviation in the phase of arrival of a symbol-carrying
signal is invisible to the Turing-machine logic of an agent, because any
physical implementation of that logic must tolerate variations within certain
limits in the phase of arrival of a symbol-carrying signal.  Thus logical
synchronization that depends on responding to phases requires extra physical
circuitry not encompassed in Turing's mathematics.

Because of the use of feedback to maintain logical synchronization
among the symbol-handling agents of an experiments, the lag in
responding to measured deviations limits physically possible behavior.
For this reason, it can be important to respond rapidly to the
physical phase at which a signal carrying a symbol arrives within a
computational cycle of an agent, which precludes waiting for an analog
to digital conversion; the agent has to respond to the phase of an
arriving signal energy in relation to a clock pulse, and the phase to
which the agent responds in immediate, physical, and {\em not} a
number.

Another way to see the non-numerical aspect of the phase of arrival of
a signal is to examine the fan-out of a symbol-carrying signal from
one agent to two other agents, and to contrast the behavior of
phases with that of symbols in this fan-out.  Apart from occasional
device failures, the receiving agents fed by a fan-out agree about
what symbol they receive.  In contrast , once the phases of arrival of
the symbol-carrying signal has been clocked by an agent and expressed in
numbers (e.g. analog to digital conversion), the clocks of the two
receiving agents will generally disagree about those phases.  We call
such expectation of unpredictable disagreement {\em idiosyncrasy}.
Idiosyncrasy due to jitter in phase arrivals is intrinsic to logical
synchronization.

The need for agents to react to idiosyncratic physical phases prior to their
expressions in numerals has an intriguing implication: the behavior of
symbols that makes makes mathematical calculations dependable depends on the
handling of unpredictable phases by symbol-handling agents.  This
interdependence between numerical symbols and non-numerical phases seems to
be a largely or completely overlooked feature of the relation between
mathematics and physics.

\section{Reversibility and irreversibility in physics: the impact of accepting agents}\label{sec:3}
The basic equations of physics involve a global time coordinate $t$ and are
invariant under the transformation $t \rightarrow -t$. This invariance
appears at first sight to impose time reversibility as a principle of
physics, in conflict with the phenomenological equations postulated in
thermodynamics and statistical mechanics that assert, as amply confirmed by
experiment, that heat flows from hot to cold and not the other way around.
We suggest that the inference that the invariance in the equations implies a
corresponding invariance in the physical world comes from a traditional way
of viewing the relation between equations on the blackboard and physical
behavior on the workbench.  One thinks of this relation in terms of
\emph{error}.  The idea is that numbers from theory differ from numbers
expressing evidence by some ``error,'' hopefully small.  If one assumes that
the relation between explanatory equations and the evidence explained is
encompassed by the notion of error, the invariance of equations involving a
time variable $t$ under the transformation $t \rightarrow -t$ indeed implies
time reversibility as a principle of physics.

But if, as discussed above, one accepts unpredictability in symbol-handling
agents necessary to the linking of the blackboard of theory with the
workbench of experiment, there is long-overlooked room for irreversibility in
physics. To see this source of irreversibility, consider the symbol-handling
agent functioning as a process-control computer. Whatever can be computed
from fixed inputs can be computed logically reversibly, as follows from
Bennett's introduction of a logically reversible Turing machine
\cite{bennett73}.  Logical reversibility, however, cannot work for a
symbol-handling agent because, being a choice machine, its inputs are neither
fixed nor predictable: one cannot ``back compute'' how some unpredictable
``external operator'' makes its choice of symbol to write on the scanned
square of the symbol-handling agent.  The agent responds to symbols put into
it unpredictably, so that an attempt to run a program backwards encounters
needs to determine what generated an unpredictable symbol.  The need cannot
be met, from which we conclude that the operation of an agent is
intrinsically irreversible, thereby introducing an overlooked source of
irreversibility into physics.

Given that $t$ as a variable global time coordinate is much used in
expressing motion, there is still the question: is this necessary; can one
formulate dynamics without a global $t$-variable?  Indeed one can, albeit at
a certain cost.  As described in [\citenum{aop14,qip15}], one can locate
happenings near the symbol-handling agents of a network by introducing a
``local'' variable $t_A$ for the clock reading of (adjustable) clock of the
agent $A$ of the network.  As described in the cited references, one needs to
attribute to each agent's clock an unadjusted (possibly idiosyncratic)
reference.  That is, one views the adjustable clock as moving a clock hand
linked through an adjustable ``gear box'' to an unadjusted reference clock
hand.  One expresses the unadjusted hand by a variable $t_{A,0}$, related to
the adjusted hand by a function $t_{A,0}(t_A)$. One also needs functions that
express signal propagation.  To this end one introduces the reading of the
adjustable clock of an agent $B$ at the receipt of a signal transmitted from
agent $A$ at a reading $t_A$ of $A$'s adjustable clock.  This is expressed in
[\citenum{aop14,qip15}] by a relation denoted $\abr$.  These functions define
graphs having vertices labeled by readings of agents' clocks.  These graphs
display communications among the symbol-handling agents of a network, and the
graphs offer a reference system for locating events close to ticks of the
clocks of agents, a basis that assumes no spacetime manifold, let alone a
metric tensor field.  In particular such a reference system serves to express
evidence of the readings of physical clocks that can be used to support or to
refute one or another assumption of a spacetime manifold with an assigned
metric tensor field \cite{aop14}.

If one invokes the optional hypothesis of spacetime manifold with a metric by
which to determine signal propagation, one can express the functions listed
above as functions of spacetime coordinates $x$, e.g. $t_A(x)$,
$t_{A,0,}(x)$, etc., thereby achieving a much more economical representation
of location, but then one loses the structure of clock readings related by
signal propagations that allows for testing the hypothesis.  And in some
cases, spacetime coordinates are unavailable or fail to supply a basis for
location that is available from a network of symbol-handling agents
\cite{aop14,qip15}.

\section{Discussion}\label{sec:4}
\subsection{Objections to the notion of an ``agent''}
Why would theoretical physicists so often think with the relation between the
blackboard and the workbench as if error were all that mattered?  For one
thing, one often needs to attend to errors as gaps between numbers predicted in
theory and numbers reported from experiments. We suggest, however, that this
way of thinking comes from a tradition established with Descartes and
reinforced by Hume, a tradition of rejecting any concept of agent or agency in
theoretical physics.  For that reason introducing a concept of agents, in
particular the symbol-handling agents introduced above, has a residual
antagonism to overcome.

In the Merriam-Webster dictionary we find the following relevant definition
of `agent'.
\begin{quote}
  \textbf{agent}: something that produces or is capable of producing a
  certain effect; an active or efficient cause.
\end{quote}
Because of the association of ``agent'' with ``cause,'' a distaste for
``cause'' rubs off on ``agent''.  Traditionally in physics one avoids using or
even discussing notions of ``cause'' and ``agency.''  Although not much
discussed in physics, these notions are discussed by philosophers and
historians, to which we now refer.  The notion of an agent as an ``active
cause'' runs into a logical swamp announced in Hume's demonstration that
causality is unprovable \cite{hume}, to which Einstein agrees
\cite{weltbild}.  Russell \cite{russell}, as did Locke before him, says that
the conception of cause likely comes from volition, i.e. from ones experience
of doing something that one \emph{wants to do}, and is applied to other
agents by analogy; for Hume `cause' is a feeling or intuition stemming from
habitual association.  As Russell puts it in discussing Hume's objections to
the conception of cause,
\begin{quote}
  I think perhaps the strongest argument on Hume's side is to be derived from
  the character of causal laws in physics. It appears that simple rules of
  the form ``A causes B'' are never to be admitted in science, except as
  crude suggestions in early stages. The causal laws by which such simple
  rules are replaced in well-developed sciences are so complex that no one
  can suppose them given in perception; they are all, obviously, elaborate
  inferences from the observed course of nature. I am leaving out of account
  modern quantum theory, which reinforces the above conclusion. So far as the
  physical sciences are concerned, Hume is wholly in the right: such
  propositions as ``A causes B'' are never to be accepted, and our
  inclination to accept them is to be explained by the laws of habit and
  association. 
\end{quote}

In her recent book, ``The Restless Clock,'' J. Riskin starts a discussion of
agency with: ``By `agency' I mean \ldots a capacity to \ldots do things in a way
that is neither predetermined nor random. Its opposite is passivity.'' In the
next paragraph she writes:
\begin{quote}
\ldots, the scientific principle banning ascriptions of agency to natural
  things supposes a material world that is essentially passive.  This principle
  came into dominion around the middle of the seventeenth century, during the
  period that historians generally identify as the origin moment of modern
  science, or the New Science as its inventors called it.  It is the informing
  axiom of a mechanistic approach to science. Mechanism, the core paradigm of
  modern science from the mid-seventeenth century onward, describes the world
  as a machine---a great clock, in seventeenth- and eighteenth-century
  imagery---whose parts are made of inert matter, moving only when set in
  motion by some external force, such as a clockmaker winding the spring.
  According to this originally seventeenth-century model, a mechanism is
  something lacking agency, produced and moved by outside forces; and nature,
  as a great mechanism, is similarly passive.  Assuming that living beings are
  part of nature, according to this model, they too must be rationally explicable
  without appeal to intentions or desires, agency or will [\citenum{riskin}], p.3.
\end{quote}

\subsection{Rehabilitating a ``lean'' agency in physics, based on recognizing unpredictability}
Opening a breach against this resistance, authors of papers on quantum
information science habitually speak of communications between agents Alice
and Bob, sometimes subject to eavesdropping by Eve and Charlie, etc.;
however, these authors speak in an offhand way, as a convenience, without
claiming that agents are necessary to the physics described, nor do they
touch on logical synchronization.  Similarly, synthetic biology papers deal
with living cells as computational agents responding to unpredictable inputs,
but specify the agents only vaguely, without facing the question whether
speaking of agency is a minor convenience or the introduction of a
significant novelty.  We attribute the announcement of occurrences of quantum
outcomes to symbol-handling agents, leading of course to one aspect of the
introduction of irreversibility.  This aspect of irreversibility was
introduced over twenty years ago as `quantum jumps' that, that admit the
unpredictability of measurements into physics, but without explicit mention
of any notion of agents, and again without pointing to the logical
synchronization needed for the functioning of symbol-handling
agents\cite{jump}.
 
Giving agency a fundamental place in physics would be a big change.  We
propose that the proof of a gap between evidence and its explanations that
has to be bridged by a guess is a sufficient reason to at least re-open the
discussion of agency. The model outlined above of symbol-handling agents and
their dependence on logical synchronization adds force to this proposal.  It
is worth noting that the introduction of unpredictability is free of any need
for `volition', let along `values', on the part of a symbol-handling agent.
For example, a symbol-handling agent fed an input from an unpredictable
photo-detector evokes no need to speak of volition, nor do we need to speak
of such an agent as ``wanting to do something''; recognizing its
unpredictability suffices to assure its ``agency'' in the leaned-down sense
that we have explicated in the sections above.

The recognition of the physical reference systems as constructions involving
symbol-handling agents with their adjustable clocks brings into a modern
technical context the contention of Saint Augustine of time as a
construction.  As Russell quotes a translation of the Confessions of Saint
Augustine, ``Time was created when the world was created.''
[\citenum{russell}] (p. 353).  As we argued in [\citenum{aop14,qip15}], the
``time'' relevant to physics is created not just when the world was created,
but is constructed to serve particular purposes. For many purposes, the time
disseminated by the National Institutes of Science and Technology (NIST) or
by other standards organizations is convenient and adequate; however, some
investigations, including some yet to be designed, require, in place of ``NIST
time,'' a special-purpose network of symbol-handling agents logically
synchronized to one another.

Acceptance of the construction of networks of logically synchronized
symbol-handling agents to serve as reference frames for particular
investigations would shift the ground on which physics stands,
by doing away with ``time'' as an indispensable externality.
A partial recognition of the construction needed to produce ``time'' is
found among those who develop and implement international
time standards.  For example, David Allan wrote:
\begin{quote} 
The fact is that time as we now generate it is dependent upon defined
origins, a defined resonance in the cesium atom, interrogating
electronics, induced biases, timescale algorithms, and random
perturbations from the ideal. Hence, at a significant level, time---as
man generates it by the best means available to him---is an
artifact. Corollaries to this are that every clock disagrees with
every other clock essentially always, and no clock keeps ideal or
``true'' time in an abstract sense except as we may choose to define
it. \cite{allan87}
\end{quote}
In spite of this known role of people in generating the ``NIST time,'' it
seems to us that physics is haunted by the ghost of old sense of ``time'' (or
in a relativistic setting, spacetime) as an invisible ``ether'' in which we
all live but can't see.  This ghost is inherited from Descartes and from
Newtonian physics---the idea that our bodies and all that we investigate
physically are made of bits and pieces that passively respond ``in time'' to
forces, with no room in physics for our own agency.  We contend that although
the concept of ``ether'' was discredited by special and general relativity,
it still exerts its mythical power, distracting investigators from
opportunities to take responsibility for generating the reference systems for
their particular investigations, e.g.  by designing and implementing networks
of symbol-handling agents.  Accepting the primacy of locally constructed,
special-purpose networks of logically synchronized symbol-handling agents to
serve as reference frames for particular investigations would shift the
ground on which physics stands, by doing away with ``time'' as an
indispensable externality.

Reference frames provided by networks of symbol-handling agents, with their
feedback loops that respond to unpredictable events in order to to try to
maintain logical synchronization, constitute tools permitting a range of
experimental investigations.  Such tools have been used for a long time, and,
for example, are used in the Laser Interferometer Gravitational-Wave
Observatory (LIGO).

In conclusion, we recognize that:
\begin{enumerate}
\item something announces unpredictable outcomes of measurements; and
  \item something puts unpredictable equations on the blackboard.
\end{enumerate}
These two propositions raise the question for theoretical physics: Do we as
physicists want to ignore this something?  Can we overcome our prejudice
against agency in physical explanations to extend the reach of basic physics
beyond time-reversible equations to encompass new sources of irreversibility?
We close with a last question: How may recognizing the dependence of symbolic
communication on physical, non-numeric phase management impact collaboration
between mathematicians and experimental physicists?

\section*{Acknowledgment}
Our thanks to  Jessica Riskin for an exchange of emails concerning the
historical pulling and tugging over the role in physics, or lack of it,
for an agent.

\end{document}